\documentclass{article}
\usepackage{graphicx} 
\usepackage{amssymb}
\usepackage{amsmath}
\usepackage{booktabs}
\usepackage{adjustbox}
\usepackage[square, numbers]{natbib}
\usepackage{algorithm}
\usepackage{algorithmic}

\usepackage[table,xcdraw]{xcolor}
\usepackage[inkscapelatex=false]{svg}
\usepackage[a4paper, total={6in, 8in}]{geometry}
\usepackage{multirow}
\usepackage{subcaption}
\usepackage{url}

\usepackage{authblk}

\title{NEBULA: A National Scale Dataset for Neighbourhood-Level Urban Building Energy Modelling for England and Wales}
\author[1]{Grace Colverd}
\author[1,2]{Ronita Bardhan}
\author[1]{Jonathan Cullen}
\affil[1]{Department of Engineering, University of Cambridge }
\affil[2]{Sustainable Design Group, Department of Architecture, University of Cambridge}

\date{Pre-print, submitted to Scientific Data November 2024}
\begin{document}
\maketitle

\begin{abstract}
Buildings are significant contributors to global greenhouse gas emissions, accounting for 26\% of global energy sector emissions in 2022. Meeting net zero goals requires a rapid reduction in building emissions, both directly from the buildings and indirectly from the production of electricity and heat used in buildings. National energy planning for net zero demands both detailed and comprehensive building energy consumption data. However, geo-located building-level energy data is rarely available in Europe, with analysis typically relying on anonymised, simulated or low-resolution data. To address this problem, we introduce a dataset of \textbf{N}eighbourhood \textbf{E}nergy, \textbf{B}uildings, and \textbf{U}rban \textbf{La}ndscapes (NEBULA) for modelling domestic energy consumption for small neighbourhoods (5-150 households). NEBULA integrates data on building characteristics, climate, urbanisation, environment, and socio-demographics and contains 609,964 samples across England and Wales.

\end{abstract}

\section{Background and Summary}\label{intro_sec}

Urban building energy modelling (UBEM) is an important tool for understanding building stocks and their associated energy use. Modern statistical UBEM methods rely on large samples of building-level energy data and associated explanatory building and occupant factors (Ali et al. \cite{ali_review_2021}). The energy data is typically collected from utility companies, or published by local governments. Privacy concerns around building-level energy data and differing global regulations have led to a lack of open-access building-level energy datasets, as noted by Jin et. al \cite{jin_review_2023}. The majority of open energy datasets are located in cities in the US, 22 out of a surveyed 33 \cite{jin_review_2023}. This concentration of data raises concerns about the potential overfitting of forecasting models to these open datasets (Kazmi et al. \cite{kazmi_ten_2023}), and limits the national planning of decarbonising building stocks. 
Of the open datasets reviewed, two datasets were from the UK. The first is the National Energy Efficiency Data Framework (NEED), an anonymised dataset released with representative data on buildings within the UK. The full NEED dataset contains 4 million samples containing building and occupant characteristics and energy consumption \cite{department_for_energy_security_and_net_zero_national_2023}. Anonymised building datasets like NEED do not enable fine-grained spatial analysis of energy consumption. Buildings within NEED are tagged with their region, but localised analysis is not possible.  
The second is the Energy Performance of Buildings Data which provides access to the Energy Performance Certificates (EPCs) of buildings in England and Wales. EPCs are a standardized method of assessing the energy efficiency and environmental performance of a building. These certificates, mandated by the European Union in 2003, provide quantifiable data regarding a dwelling's energy consumption and emissions \cite{crawley_quantifying_2019}. The European Parliament mandate on EPCs allowed them to be calculated either through direct measurement of energy use or via methodological calculation \cite{european_parliament_2010}. EPCs have been mandatory for any house sold or rented in the UK Union since 2008 (DESNZ \cite{department_for_energy_2021}). EPCs suffer from a noted performance gap where actual energy consumption differs from predictions (De Wilde \cite{de_wilde_gap_2014}). Few et al. \cite{few_over-prediction_2023} found that EPCS tend to over-predict primary energy intensity, with the over-prediction worsening for lower EPC bands (8\% for band C vs. 48\% for bands F and G), persisting even when matching assumptions around occupancy and heating styles. Furthermore, Jenkins et al. \cite{jenkins_investigating_2017} found that EPC ratings for the same building could vary by up to two bands across different assessments. 

Closed-source building-level energy datasets also exist in the literature. In the UK, the 3DStock model exemplifies a closed-source approach. It links building-level energy data with Energy Performance Certificates (EPCs) along with other data sources to develop 3D building stock models for various areas, including London (Evans et al. \cite{evans_3dstock_2016}, Godoy-Shimizu et al. \cite{godoy-shimizu_producing_2024}). The potential privacy concerns of the building energy data mean that access to this model is tightly controlled. 

The NEBULA (\textbf{N}eighbourhood \textbf{E}nergy, \textbf{B}uildings, and \textbf{U}rban \textbf{L}andscapes) dataset presented in this paper offers an alternative: a novel, privacy-preserving, geo-located, open source dataset for energy modelling in the UK at the neighbourhood level (5-150 households). We include domestic energy data for 2022, avoiding the challenges of simulated data or EPC unreliability. Working at the postcode level avoids privacy concerns and enables open-source publication. Our methodology is a true bottom-up methodology, deriving postcode-level attributes relating to the building stock, climate, urbanisation and socio-demographics. The data has already been used to provide energy benchmarks for cities in the UK (Colverd et al. \cite{colverd_benchmarking_2024}) and is well set up for predictive modelling at the postcode level.

\section{Methods}
The following section describes the methodology for the dataset generation and the processes taken to derive the final dataset. 

\subsection{Dataset Generation}
The NEBULA dataset currently includes seven themes of variables: building stock, building typology, building age, region, urbanisation, climate and socio-demographics. These variables are derived from six datasets, open-source or accessible under academic licenses. Table \ref{tab:data-sources} lists the input data sources and their licences. In total, NEBULA contains 242 variables (124 of which relate to socio-demographics, 110 to buildings, energy and environment and 11 to regional variables) and is in tabular format indexed by Postcode. The full list of variables and metadata (Column Name, Data Type, Category, Direct/Derived, Source, Description and Usage Notes) is provided in supplementary information due to its size. A selection of the variables across the themes is given in 
Tables \ref{tab:nb_1} and \ref{tab:nb_2}. The following sections describe the variable creation process for each theme using a single postcode \textit{PC} as an example.

\begin{table}[]
\centering
\caption{Selection of NEBULA Variable List Part 1. The count of variables in each theme is given in (X) in name. The full variable list and metadata are given in Supplementary Information. `Clean' refers to residential buildings with residential typologies (excluding outbuildings and unknowns).  }
\label{tab:nb_1}
\begin{tabular}{@{}ll@{}}
\toprule
\textbf{Building Stock (37)}   & \textbf{Typology (34)}                    \\ \midrule
Count of Buildings            & Pct 2 Storeys Terraces w/ Rear Extension \\
Total Ground Floor Area       & Pct 3-4 Storey and Smaller Flats         \\
Total Heated Floor Area (FC)  & Pct Domestic Outbuilding                 \\
Total Heated Floor Area (H)   & Pct Large Detached                       \\
Total Basement Floor Area     & Pct Large Semi-Detached                  \\
Count Listed Builds           & Pct Linked and Step Linked Premises      \\
Count Domestic Outbuildings   & Pct Medium Height Flats 5-6 Storeys      \\
Total Outbuilding Floor Area  & Pct Planned Balanced Mixed Estates       \\
Count Clean Builds            & Pct Semi Type House in Multiples         \\
Total Clean Ground Floor Area & Pct Small Low Terraces                   \\
Total Clean Heated Area (FC)  & Pct Standard Size Detached               \\
Total Clean Heated Area (H)   & Pct Standard Size Semi Detached          \\
Pct of Clean Builds           & Pct Tall Flats 6-15 Storeys              \\
Pct of Clean Res. Builds      & Pct Tall Terraces 3-4 Storeys            \\
Pct Listed Buildings                   & Pct Very Large Detached                  \\
\textbf{}                     & Pct Very Tall Point Block Flats          \\
                              & Pct All Unknown Typology                         \\ \midrule
\textbf{Age (16)}              & \textbf{Weather (6)}                     \\ \midrule
Pct 1919-1944                 & HDD                                      \\
Pct 1945-1959                 & CDD                                      \\
Pct 1960-1979                 & HDD Summer                               \\
Pct 1980-1989                 & CDD Summer                               \\
Pct 1990-1999                 & HDD Winter                               \\
Pct Post 1999                 & CDD Winter                               \\
Pct Pre 1919                  & \textbf{}                                \\
Pct Unknown Age               &                                          \\ \bottomrule
\end{tabular}
\end{table}

\begin{table}[]
\centering 
\caption{Nebula Variable List Part 2.}
\label{tab:nb_2}

\begin{tabular}{@{}ll@{}}
\toprule
\textbf{Region (11)}               & \textbf{Energy (11)}        \\ \midrule
Region                            & Annual Gas / Electricity   \\
Output Area                       & Mean Gas / Electricity     \\
LSOA                              & Median Gas / Electricity   \\
MSOA                              & Count of Meters (Gas)      \\
Local Authority                   & Count of Meters (Elec)     \\ \midrule
\textbf{Urbanisation (3)}         &                            \\ \midrule
Postcode Area                     & Rural/Urban classification \\
Postcode Housing Density          &                            \\ \midrule
\multicolumn{2}{c}{\textbf{Socio-Demographics (124)}}            \\ \midrule
Ethnicity (20)                    & Household Size (9)         \\
Economic Activity (20)            & Occupation (10)            \\
Household Composition (6)         & Central Heating (13)       \\
Tenure (9)                        & Sex (2)                    \\
Occupancy Rating (6)              & Deprivation (6)            \\
Deprivation (6)                   & Highest Qualification (8)  \\
Socioeconomic Classification (10) & Bedroom Count (5)          \\ \bottomrule
\end{tabular}
\end{table}

\begin{table}[]
\caption{NEBULA Data Sources. ONS: Office for National Statistics. DESNZ: Department for Energy Security and Net Zero. OS: Ordnance Survey, © Crown copyright and database rights 2024 Ordnance Survey (AC0000851941). UK Open Government Licence is a worldwide, royalty-free, perpetual, non-exclusive licence. Education licence refers to free access for those in non-commercial research or educational settings.
}
\label{tab:data-sources}
\centering
\begin{tabular}{@{}lllll@{}}
\toprule
Dataset        & Frequency & Year  & Ref & License \\ \midrule
Building Stock & Annual    & 2022       & Verisk \cite{the_geoinformation_group_limited_digital_nodate} &  Educational  \\
Temperature    & Monthly   & 2022       & Met Office \cite{uk_met_office_haduk} & Open Government v3.0 \\
UK Regions    & Annual    & 2021       & ONS \cite{ons_local_2023} & Open Government v3.0  \\
Energy         & Annual    & 2022       & DESNZ \cite{desnz_electricity_2024} &  Open Government v3.0\\
Postcode       &   Quarterly        &    2022        & OS 
\cite{os_codepoint} & Educational   \\
Census         & Decade     & 2021       & ONS \cite{ons_census_2021} &  Open Government v3.0  \\
Urban/Rural classification & N/A & 2011 & ONS \cite{office_for_national_statistics_2011} & Open Government v3.0  \\\bottomrule
\end{tabular}

\end{table}

\subsubsection{Energy}
The energy data within NEBULA is drawn from the records published by the UK Department for Energy Security and Net Zero (DESNZ \cite{desnz_electricity_2024}). This data details annual domestic energy consumption, with variables including the count of meters per postcode and total, median and mean consumption per postcode, for both gas and electricity. DESNZ provides energy consumption at the postcode level for postcodes with at least five meters, and gas data is regionally adjusted for the temperature to allow for inter-year comparison. Here we use energy data from 2022. 
 
\subsubsection{Building Stock}
The building stock dataset used in this work is the Verisk UK Buildings dataset (The GeoInformation Group Ltd), published by the Edina Digimap service \cite{the_geoinformation_group_limited_digital_nodate}. The Verisk data includes key attributes such as building footprint, height, age and typology for all buildings in the UK. We chose this building stock data over open source alternative due to the greater accuracy of building footprints as noted in  Krapf et al. \cite{krapf_points_2023}. Verisk have confirmed the publication of the NEBULA dataset as acceptable for open source, given building level data is not included.

Before deriving postcode attributes, we implemented cleaning and pre-processing stages. Building heights and ages were cleaned and categorized into discrete groups. Various errors exist within the dataset. We focused on reducing those in floor count and height based on thresholds derived from average building parameters. We first calculate the average floor height $F_H$:
\begin{equation}
    F_H = \frac{H}{F}
\end{equation}
where $H$ is building height and $F$ is building floor count. We then calculate the minimum width of the building footprint $W_M$.

We validate the height and floor count for each building using the following algorithms:
\begin{equation}
\text{if } H \geq 3 \cdot W_M \text{ OR } ( H < 2 \text{ AND numeric(FC) )} \implies H = \text{NULL}
\end{equation}
\begin{equation}
    \text{if } (F <= 2.3 \text{ OR } F > 5.3 ) \text{ AND } H < 3\cdot W_M  \implies F= \text{NULL}
\end{equation}

Height is set to invalid if height is greater than three times the $W_M$, or if height is less than 2m and there is a valid floor count present. Floor count is set to invalid if outside the thresholds for average floor height and if height is less than 3x $W_M$. These thresholds are loosely based on building standards for average ceiling height (2.3-2.5m) (Coates \cite{coates_whats_2024}), including a larger upper variation allowing for larger historic houses and varying roof heights. Missing or null values for $H$ and $F$ are then filled using local averages from within the postcode. We then calculate the global average floor count ($F_G$) for each building. The global average tables give an average floor count for a building height/age combination, generated from all buildings in the UK Buildings dataset.

We use a dual approach to calculate building floor area per building, using both floor count (F) and global average floor count derived from height ($F_H$):
\begin{equation}
    A_H = F_{H} * F_A
\end{equation}
\begin{equation}
    A_{F} = F * F_A
\end{equation}
where $F_A$ is the building footprint area and $F_H$ is the global average floor count for $H$, $F$ is the floor count. This method takes advantage of all the data available and mitigates against errors in either height or floor count. 

In cases where height and floor count data align closely, these two derived values converge, whereas discrepancies between them indicate potential data quality issues. A `confidence floor area' metric was developed to quantify data reliability, calculated from the magnitude of divergence between these two values. The thresholds for the confidence metric area:
\begin{itemize}
    \item High Confidence:  $x \leq$ 3\% difference 
    \item Medium: $3 < x \leq10 $\% difference 
    \item Low: $10 < x \leq 25$\% difference 
    \item Very Low: $>$ 25\% difference 
\end{itemize}

\begin{figure}
      \caption{Example of the dual method for matching buildings to postcode Visualisation code adapted from Lipson \cite{lipson_plotting_2021}. Satellite image downloaded from 
\cite{openstreetmap_contributors__planet_2017}. Footprints via Digital Map Data © Verisk (The GeoInformation Group Ltd.) \cite{the_geoinformation_group_limited_digital_nodate} }
     \label{fig:pc_example}
     \centering
     \includegraphics[width=0.6\linewidth]{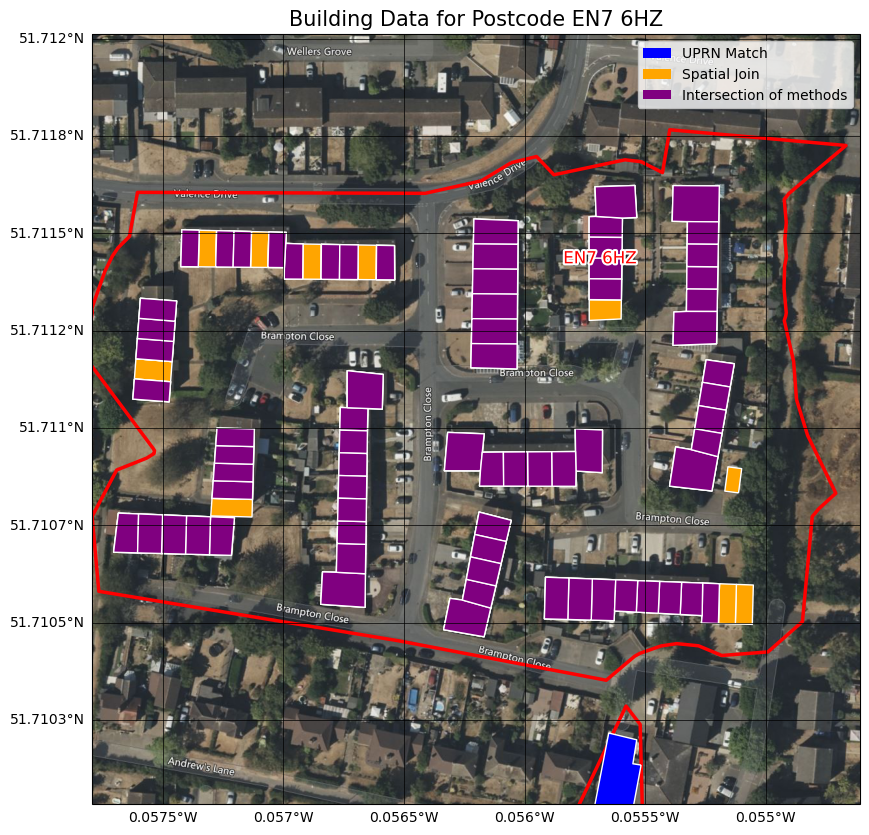}

 \end{figure}

After dataset pre-processing, all buildings within \textit{PC} are identified, using a spatial join between building footprints and postcode polygons and a UPRN to Postcode mapping. This identifies both dwellings and outbuildings (with no UPRNs). Figure \ref{fig:pc_example} shows an example of this matching. The cleaned building stock for \textit{PC} is then aggregated into small neighbourhood-level variables. Variables are either count, total sum or percentage e.g. count of residential buildings, total floor area for residential buildings, percentage of residential builds in postcode. For the derived variables $A_H$ and $A_F$, we also log the count of nulls within the base metric. 

\subsubsection{Climate}
Heating and Cooling Degree days are calculated from monthly Temperature data using Equations \ref{eqn1} and \ref{equ2}. We include temperature data for 2022 from HADUK \cite{uk_met_office_haduk}. The closest temperature sensor to the postcode is identified, and monthly values for HDD and CDD are calculated using the set points of 18 and 15.5. 

\begin{equation}\label{eqn1}
    \text{HDD} = Min(15.5 - \text{Temp}, 0)
\end{equation}
\begin{equation}\label{equ2}
    \text{CDD} = Min(\text{Temp} - 18, 0)
\end{equation}
Pre-processing of temperature data includes spatial interpolating of nulls in the x and y direction with a limit of 10km (limit of 10 neighbour fills). HDD and CDD are calculated for summer, winter, and annually for 2022, using six-month windows. 
   \begin{equation}
       \text{summer months} = [4, 5, 6, 7, 8, 9]    \end{equation}
       \begin{equation}
       \text{winter months} = [1, 2, 3, 10, 11, 12]
   \end{equation}

\subsubsection{Urbanisation}
Urbanisation is included in both the area of the postcode and the housing density.  NEBULA uses postcode shapefiles to calculate the area of the postcode $P_A$, using the Python library GeoPandas \cite{jordahl_geopandasgeopandas_2020}. Postcode density $P_d$ is calculated:
\begin{equation}\label{urb_density}
     P_d = \frac{\sum_i{F_{A_i}}}{P_A}
\end{equation} 
where $F_{A_i}$ is the building footprint area. We also include the 2011 Urban/Rural classification data here from the UK Government, as tagged in Table \ref{tab:data-sources}.   

\subsubsection{Regions}
NEBULA uses a mapping from the Office of National Statistics (ONS) to match \textit{PC} to Output Area (OA), and subsequent higher geographic census regions (Lower Super Output Area (LSOA), Middle Super Output Area (MSOA) and Local Authority (LA)). We use the 2021 mappings to align with the 2021 census geographical boundaries. 

\subsubsection{Socio-Demographics}
NEBULA uses the 2021 census for England and Wales to calculate socio-demographic data per \textit{PC} (ONS \cite{ons_census_2021}). The raw census data (e.g. count of households with property X) is transformed into percentage attributes at the OA level and then matched to \textit{PC} the aforementioned ONS. The 13 themes of demographics are given in table \ref{tab:nb_2}. The census themes were chosen to cover a broad range of socio-demographic factors and include highly predictive attributes identified in prior works. The total number of demographic attributes is 124.

\subsubsection{Energy Use Intensity}
Energy Use Intensity (EUI) was calculated as annual energy consumption per heated floor area. We use this to help identify outlier postcodes. We calculate the total heated floor area per PC using all residential buildings' floor area (excluding outbuildings). 
\begin{equation}
    \text{EUI} = \frac{E}{\sum_\text{PC}{A_H}}
\end{equation}
where $E$ is the total annual domestic energy consumption (gas or electric), $A_H$ is the total building area derived from height, summed over the postcode PC. 

\subsection{Final Dataset}
After attribute generation, several additional checks and balances were applied to ensure a clean dataset.

\paragraph{Household filter} We filtered based on the alignment of the number of gas meters and the count of UPRNs in a postcode. UPRNs refer to addresses and work as a proxy for the count of households. We filter out any postcodes where the difference between the count of UPRNs and the count of gas meters was greater than 40\%. 

\paragraph{EUI thresholds} Whilst energy analyses typically exclude based on total energy use, here we follow \cite{godoy-shimizu_producing_2024} and filter based on annual EUI thresholds (kWh/m$^2$). A gas EUI filter of $5<$ EUI $<500$ kWh/m$^2$ and an electricity filter of $0<$ EUI $< 150$ kWh/m$^2$ are applied. 

\paragraph{Other thresholds} Thresholds were applied to the count of buildings and total building volume of 1-200 and 50-20,000m$^2$ respectively. We also exclude postcodes with a percentage of unknown typologies $>25$. 

\begin{figure}
    \centering
        \caption{Stages of processing with the count of postcodes. }
    \label{fig:nebula_summary}
    \includegraphics[width=0.9\linewidth]{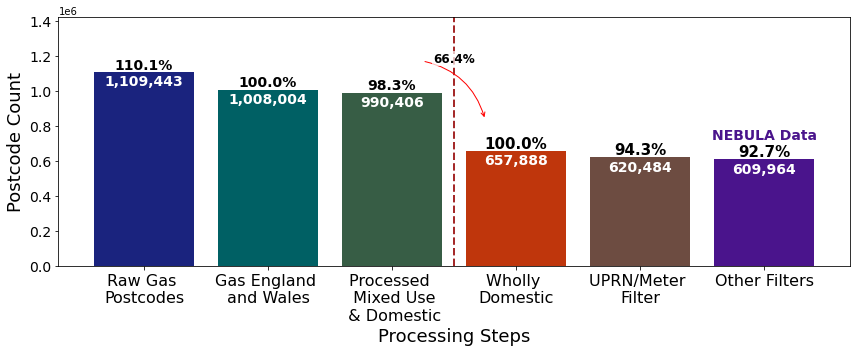}
\end{figure}

The final NEBULA dataset contains 609,964 samples. A visual representation of the processing steps and dataset sizes is given in Figure \ref{fig:nebula_summary} and the summary statistics for the dataset are given in Table \ref{tab:summary}. Our pipeline has a strong match rate when processing the original gas postcodes, with 98.3\% of postcodes matched and processed, out of gas postcodes in England and Wales (Scotland excluded on the basis of census data location). 66.4\% of these postcodes are wholly domestic based on building typology within the postcode, excluding mixed-use neighbourhoods. This was done to match the domestic energy data: commercial energy consumption is not available at the postcode level. After the domestic filter is applied, the additional filters reduce postcodes to 92.7\% of the matched domestic postcodes. This makes up the NEBULA dataset.

\begin{table}[]
\centering
\caption{Summary of final dataset, with mean, 25th, 50th and 75th percentile. PC Area: postcode area. COB = count of buildings. Fl.area = total residential building floor area.}
\label{tab:summary}
\begin{tabular}{llllllllll}
\hline
\multirow{2}{*}{Metric} & \multirow{2}{*}{\begin{tabular}[c]{@{}l@{}}PC Area\\ (m$^2$)\end{tabular}} & \multirow{2}{*}{COB} & \multirow{2}{*}{\begin{tabular}[c]{@{}l@{}}Fl.Area \\ (m$^2$)\end{tabular}} & \multicolumn{2}{l}{Total (kWh/yr)} & \multicolumn{2}{l}{Num. meters} & \multicolumn{2}{l}{EUI (kWh/yr/m$^2$)} \\ \cline{5-10} 
                        &                                                                            &                      &                                                                             & \textit{Gas}    & \textit{Elec}    & \textit{Gas}   & \textit{Elec}  & \textit{Gas}      & \textit{Elec}      \\ \hline
mean                    & 13,082                                                                     & 21                   & 2,917                                                                       & 232,188         & 64,245           & 21             & 22             & 82.4              & 23.3               \\
25\%                    & 4,330                                                                      & 9                    & 1,479                                                                       & 115,053         & 31,919           & 11             & 11             & 66.4              & 17.8               \\
50\%                    & 7,532                                                                      & 17                   & 2,453                                                                       & 194,614         & 54,181           & 18             & 18             & 80.3              & 22.4               \\
75\%                    & 13,013                                                                     & 29                   & 3,860                                                                       & 312,718         & 86,785           & 29             & 29             & 95.7              & 27.4               \\ \hline
\end{tabular}
\end{table}

\section{Data Records}
The NEBULA dataset is stored in two comma-separated value (CSV) files, indexed by postcode. The first file \path{NEBULA_englandwales_domestic_filtered.csv} contains the post-processed NEBULA data including the filters and validations given in the above sections, to offer a clean sample for energy modelling.
\path{NEBULA_englandwales_domestic_unfiltered.csv} contains the unfiltered domestic data (`Wholly domestic' in Figure \ref{fig:nebula_summary}) The full index of column names and data types is provided in a supplementary table due to its size. We provide both files to allow users to experiment with their own filters and thresholds.

\section{Technical Validation}
The pre-processing stages detailed in the methodology have corrected various errors in the underlying building stock data. Given the aforementioned prior work within the literature identifying square meters of living space as a highly influential variable influencing energy consumption, we run a sensitivity analysis on our derived variable `Total building floor area per postcode' ($Y = \sum_i F_{A_i}$, \path{`all_types_total_fl_area_H_total'}). 
Our approach, informed by Menberg et al. \cite{menberg_sensitivity_2016}, employed Morris' method for parameter screening, which systematically explores the parameter space by applying one-at-a-time variations to input parameters \cite{max_d_factorial_1991}. This method was chosen for its computational efficiency given the substantial run time of the data generation model, and the ability to identify the variation caused by each input parameter. For each of the 10 regions in England and Wales, we tested a sample of 1000 postcodes, stratified on the number of buildings per postcode to ensure representativeness. The analysis involved applying perturbations of $\pm$10\% to height and floor area, sampling from ranges of perturbations multipliers from 0.9 to 1.1 for each input.

For a set of k input parameters, Morris' method generates a sequence of k+1 points, each differing in one coordinate from the preceding one, forming a trajectory, where each point represents a model run. The magnitude of variation in model output due to a pre-defined variation of parameter i is termed the elementary effect (EE), calculated as \cite{max_d_factorial_1991, menberg_sensitivity_2016}:

\begin{equation}
    \textit{EE}_i = \frac{Y(X+e_i\Delta_i) - Y(X)}{\Delta_i}
\end{equation}
where Y is the model output, X is the parameter vector, $e_i$ is a vector of zeroes except for the i-th parameter that equals 1 and represents an incremental change in parameter i, and $\delta_i$ is the predetermined variation. EE has the same units as the model outcome Y, which for our analysis is $m^2$.

For each postcode, we run 10 Morris trajectories, examining the absolute mean ($\mu^*_i$) and standard deviation ($\sigma$) of the elementary effects:
\begin{equation}
    \mu^*_i = 0.5\sum_{t=1}^{r}|\textit{EE}_{it}|
\end{equation}

\begin{equation}
    \sigma = \sqrt{\frac{1}{(r-1)} \sum_{t=1}^{r}(\textit{EE}_{it}-\mu_i)^2}
\end{equation}
where r is the number of trajectories. 

Menberg et al. \cite{menberg_sensitivity_2016} noted that the Morris method could be unstable regarding parameter ranking in the context of building energy modelling. Still, that instability was mitigated by using the absolute median and multiple runs. Given the large number of postcodes being analysed relative to the trajectories, we calculate both the mean and median across the $\mu^*_i$ and $\sigma$ across the set of postcodes. Given the high computational cost per postcode and the range of postcodes assessed, this method efficiently compromises accuracy and computational feasibility. 

\begin{figure}
    \centering
    \begin{subfigure}{\linewidth}
        \centering
        \includegraphics[width=0.9\linewidth]{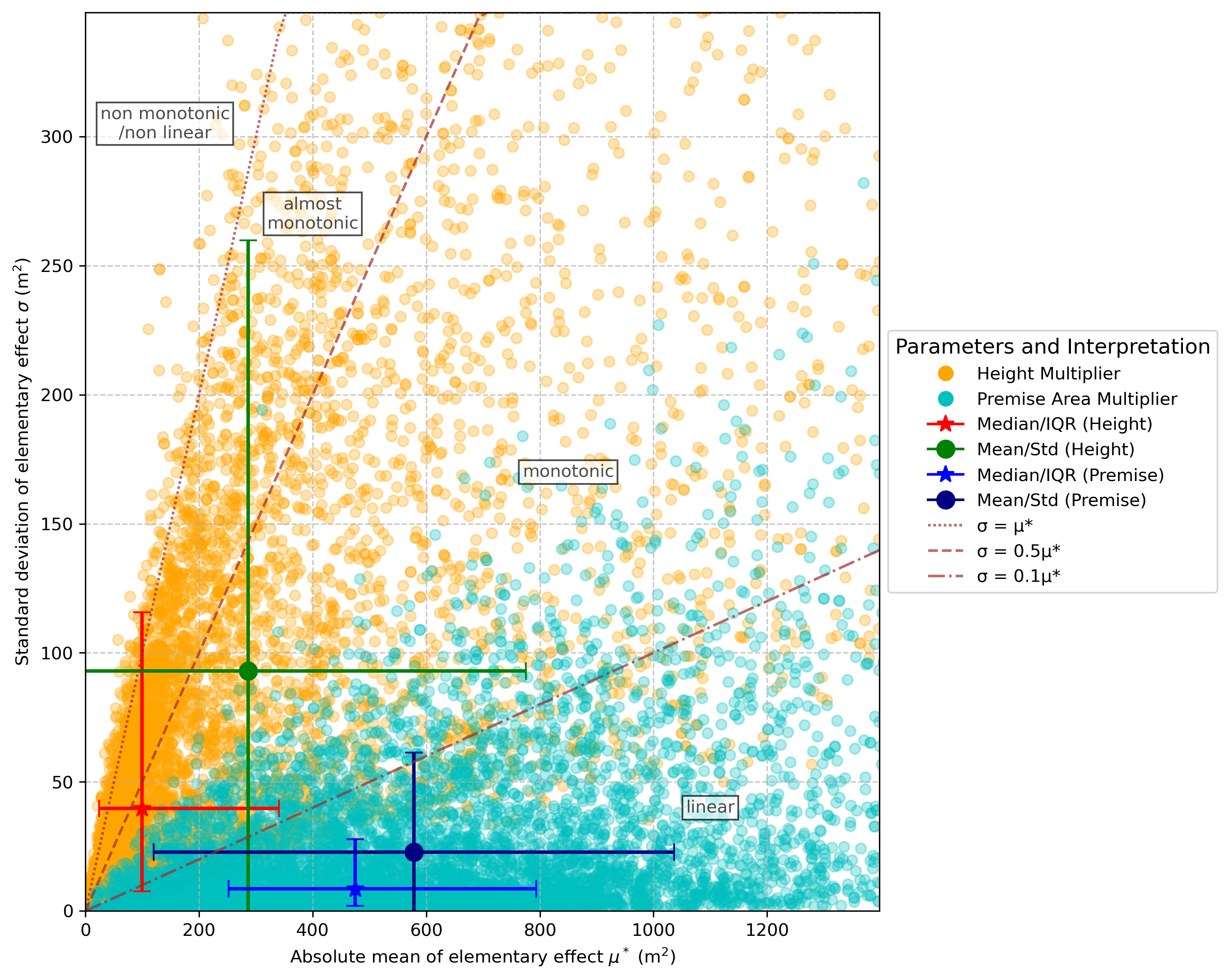}
        \caption{Results from sensitivity analysis with Morris method generating attribute `Total Building Floor Area'. Analysis run
        for 10,000 postcodes (1000 from 10 regions, stratified by building count) with 10 trajectories each assigning multipliers within $\pm$10\%. The axis is capped at the 95th quartile for visualisation purposes.}
        \label{fig:moris_method_data}
    \end{subfigure}
    
    \vspace{1em}
    
    \begin{subfigure}{\linewidth}
        \centering
        \begin{tabular}{@{}lllll@{}}
        \toprule
        \multirow{2}{*}{Variable} & \multicolumn{2}{c}{$\mu^*$ (kWh/yr)} & \multicolumn{2}{c}{$\sigma$} (kWh/yr)\\ \cmidrule(l){2-5} 
                                  & Median       & Mean         & Median        & Mean         \\ \midrule
        Height                    & 286.68       & 99.40        & 39.79         & 92.99        \\
        Premise Area              & 577.95       & 475.02       & 8.61          & 22.79        \\ \bottomrule
        \end{tabular}
        \caption{Morris parameter results}
        \label{tab:mor_params}
    \end{subfigure}
    
    \caption{Sensitivity analysis results using Morris parameter screening for generating Total building floor area per postcode.}
    \label{fig:joint_figure}
\end{figure}

\begin{figure}
    \centering
    \includegraphics[width=0.99\linewidth]{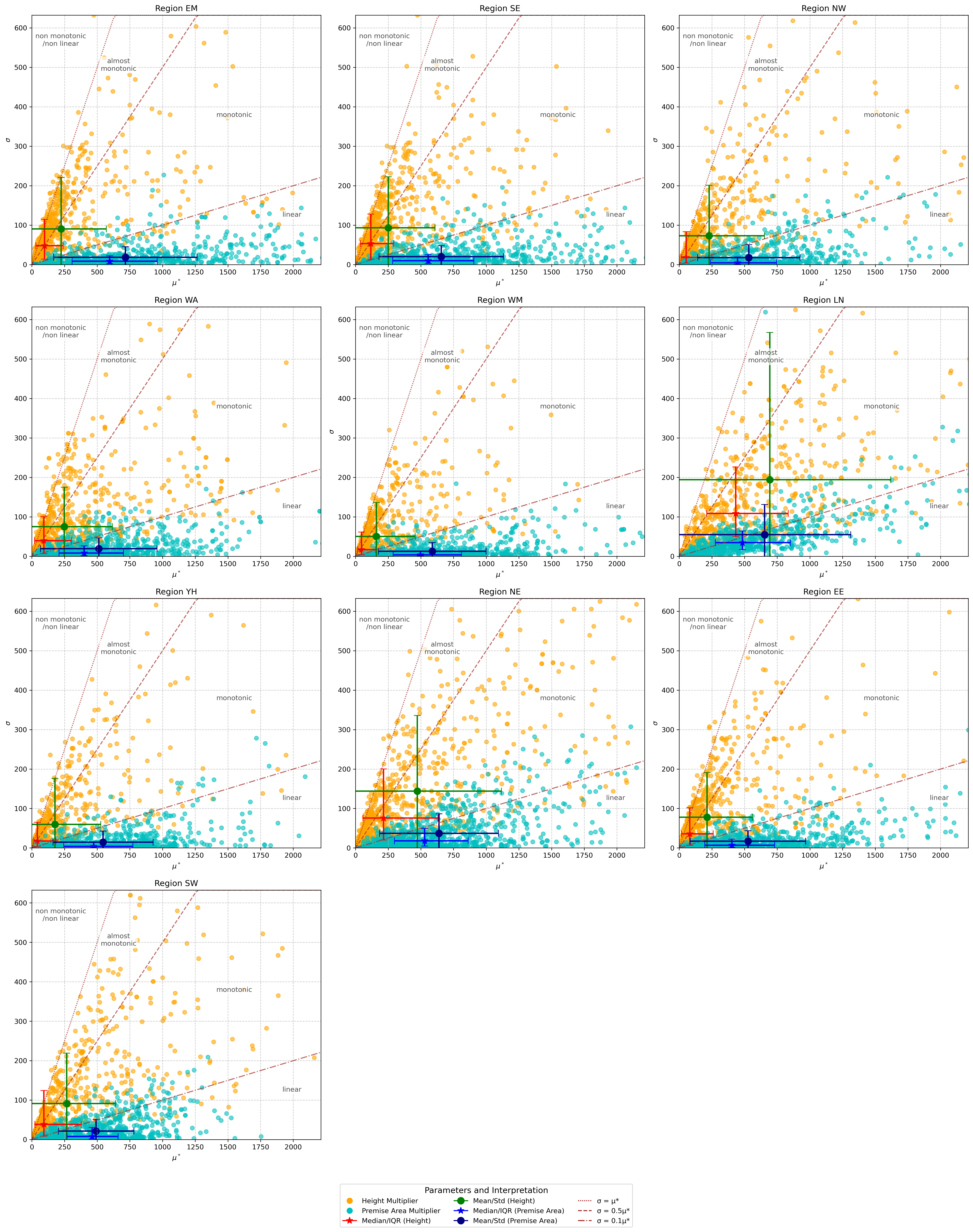}
    \caption{Regional performance for Morris parameter analysis}
    \label{fig:morris_regional}
\end{figure}

The results from evaluating 10,000 postcodes with the Morris method, with 10 trajectories are given in Figure \ref{fig:joint_figure}, which contains both the visualisation across the runs and the values for the mean and median of the Morris parameters. Following the classification scheme proposed by Garcia Sanchez et al. \cite{garcia_sanchez_application_2014} and applied by Menberg et al. \cite{menberg_sensitivity_2016}, we analysed the ratio of $\mu^*$ to $\sigma$ to characterize input parameters in terms of (non-)linearity and (non-)monotony. The results reveal that the premise area is the more influential parameter, exhibiting higher elementary effects and consequently a greater impact on Y. When examining the mean and standard deviation of the variables, there is a greater overlap between the parameters. However, the distinction is more pronounced when reviewing the median and interquartile ranges, likely due to the influence of outlier postcodes, a phenomenon also observed by Menberg et al. \cite{menberg_sensitivity_2016}. Notably, the premise area consistently falls within the linear regime, indicating a direct linear relationship between variations in the premise area and total floor area per postcode, an expected outcome given the model's algorithmic structure.
In contrast, height demonstrates monotonic to almost-monotonic behaviour, characterized by a substantially lower $\mu^*$ but a much higher $\sigma$. This pattern aligns with the model's methodology, which uses global averages across building types and height categories to derive probable floor count from an individual building's height. Consequently, this approach provides insulation against errors in building height measurements. A 10\% change in premise area equals a median 457$\pm9$m$^2$ (mean $\pm$ standard deviation) absolute change in Y (total postcode floor area), whilst a 10\% change in height equals a median 99$\pm40$m$^2$ absolute change. Set against the median Y of 2,509m$^2$ (refer to Table \ref{tab:summary}), the potential variation due to 10\% variation in premise area is 18$\pm0.3$\%, and in height is 4$\pm1.5$\%. 
The results broken down by region indicate consistent performance except for in London, which shows a greater range of Morris parameters, and a much more equal ranking to both premise area and height (refer to regional results in Figure \ref{fig:morris_regional}). 

\section{Usage Notes}
The data is provided in CSV form for ease of use. The variable's names and descriptions are all provided in the supplementary information. 

\section{Code Availability}
The code for generating the dataset is available at \url{https://github.com/graceebc9/NebulaDataset}. Repo uses Python programming language and contains details on the appropriate environment. Open source input data will be made available, whilst the user must provide the education-access datasets themselves. A Read Me offers information on running the code (locally or on a high-performance computing (HPC) machine) and the repository structure.

\section{Author Contributions}
GC conceptualized and executed the research, developed the computational methodology, conducted data preprocessing and statistical analyses, and authored the original manuscript. RB and JC provided research supervision and critical revision of the manuscript.

\section{Competing Interests}
The authors declare no competing interests. 

\section{Acknowledgements}
This work was supported by the UKRI Centre for Doctoral Training in Application of Artificial Intelligence to the study of Environmental Risks (reference EP/S022961/1).


\begin{thebibliography}{99}  

\bibitem{ali_review_2021} Ali, U. et al. Review of urban building energy modelling (UBEM) approaches methods and tools using qualitative and quantitative analysis. Energy Build. \textbf{246}, 111073 (2021).

\bibitem{jin_review_2023} \author{Jin, X., Zhang, C., Xiao, F., Li, A., Miller, C.} A review and reflection on open datasets of city-level building energy use and their applications. Energy Build. 285, 112911 (2023).

\bibitem{kazmi_ten_2023} \author{Kazmi, H., Fu, C., Miller, C.} Ten questions concerning data-driven modelling and forecasting of operational energy demand at building and urban scale. Build. Environ. 239, 110407 (2023).

\bibitem{department_for_energy_security_and_net_zero_national_2023} \author{Department for Energy Security and Net Zero} National Energy Efficiency Data-Framework (NEED) (GOV.UK, 2023).

\bibitem{crawley_quantifying_2019}
Crawley, J. et al. Quantifying the measurement error on England and Wales EPC ratings. 
Energies \textbf{12}, 3523 (2019).

\bibitem{european_parliament_2010}
European Parliament. Directive 2010/31/EU of the European Parliament and of the Council of 19 May 2010 on the energy performance of buildings. European Parliament: Brussels, Belgium (2010).

\bibitem{department_for_energy_2021}
Department for Energy Security and Net Zero \& Department for Levelling Up, Housing and Communities. Improving Energy Performance Certificates: action plan - progress report (GOV.UK, 2021).

\bibitem{de_wilde_gap_2014} \author{De Wilde, P.} The gap between predicted and measured energy performance of buildings: A framework for investigation. Autom. Constr. 41, 40-49 (2014).

\bibitem{few_over-prediction_2023} \author{Few, J., et al.} The over-prediction of energy use by EPCs in Great Britain: A comparison of EPC-modelled and metered primary energy use intensity. Energy Build. 288, 113024 (2023).

\bibitem{jenkins_investigating_2017} \author{Jenkins, D., Simpson, S., Peacock, A.} Investigating the consistency and quality of EPC ratings and assessments. Energy 138, 480-489 (2017).


\bibitem{evans_3dstock_2016} \author{Evans, S., Liddiard, R., Steadman, P.} 3DStock: A new kind of three-dimensional model of the building stock of England and Wales, for use in energy analysis. Environ. Plan. B 44 (2016).

\bibitem{godoy-shimizu_producing_2024} \author{Godoy-Shimizu, D., et al.} Producing domestic energy benchmarks using a large disaggregate stock model. Build. Serv. Eng. Res. Technol. 45, 217-239 (2024).



\bibitem{colverd_benchmarking_2024} \author{Colverd, G., Barhan, R., Cullen, J.} Benchmarking Domestic Energy Consumption using High-Resolution Neighbourhood Energy Data and City Clustering in the UK. in Proceedings of the 11th ACM International Conference on Systems for Energy-Efficient Buildings, Cities, and Transportation 121-131 (ACM, 2024).

\bibitem{desnz_electricity_2024} \author{Department for Energy Security and Net Zero} Sub-national electricity consumption data (January 2024). Available at: https://www.gov.uk/government/collections/sub-national-electricity-consumption-data (Accessed: 18 February 2024). Licensed under the Open Government Licence v.3.0.

\bibitem{the_geoinformation_group_limited_digital_nodate} \author{The GeoInformation Group Limited} Digital Map Data © The GeoInformation Group Limited 2024. Datatset: UK Buildings. Edina Digimap (2024). Available at: https://digimap.edina.ac.uk/verisk (Accessed: 27 January 2024).

\bibitem{krapf_points_2023} \author{Krapf, S., Mayer, K.,  Fischer, M.} Points for energy renovation (PointER): A point cloud dataset of a million buildings linked to energy features. Scientific Data 10, 639 (2023). doi:10.1038/s41597-023-02544-x


\bibitem{coates_whats_2024} \author{Coates, A.} What's the UK's standard ceiling height for houses, extensions and loft conversions? Designs in Detail (2024); available at https://www.designsindetail.com/articles/whats-the-uks-standard-ceiling-height-for-houses-extensions-and-loft-conversions

\bibitem{lipson_plotting_2021} \author{Lipson, M.} Plotting OpenStreetMap images with Cartopy. The Urbanist (2021); available at \url{https://www.theurbanist.com.au/2021/03/plotting-openstreetmap-images-with-cartopy/}

\bibitem{openstreetmap_contributors__planet_2017} \author{OpenStreetMap contributors} Planet dump retrieved from https://planet.osm.org (OpenStreetMap, 2017).

\bibitem{uk_met_office_haduk} \author{UK Met Office} HadUK-Grid Gridded climate observations for the UK. Available at: https://www.metoffice.gov.uk/research/climate/maps-and-data/data/haduk-grid/haduk-grid (Accessed: 18 May 2023). Licensed under the Open Government Licence v.3.0.




\bibitem{jordahl_geopandasgeopandas_2020} \author{Jordahl, K., Van den Bossche, J., Fleischmann, M. et al.} GeoPandas: v0.8.1. Zenodo (2020). doi:10.5281/zenodo.3946761


\bibitem{ons_census_2021} \author{Office for National Statistics} UK Census (2021). Available at: https://www.ons.gov.uk/census (Accessed: 15 March 2024). Licensed under the Open Government Licence v.3.0.


\bibitem{menberg_sensitivity_2016} \author{Menberg, K., Heo, Y. \& Choudhary, R.} Sensitivity analysis methods for building energy models: Comparing computational costs and extractable information. Energy Build. 133, 433-445 (2016).

\bibitem{max_d_factorial_1991} \author{Morris, M. D.} Factorial sampling plans for preliminary computational experiments. Technometrics 33, 161-174 (1991).

\bibitem{garcia_sanchez_application_2014} \author{Garcia Sanchez, D., Lacarriere, B., Musy, M., Bourges, B.} Application of sensitivity analysis in building energy simulations: Combining first- and second-order elementary effects methods. Energy Build. 68, 741-750 (2014).





\bibitem{ons_local_2023} \author{Office for National Statistics} Local Authority District to Region (April 2022) Lookup in EN. Open Geography Portal (November 2023). Available at: https://geoportal.statistics.gov.uk/ (Accessed: 8 October 2024). Licensed under the Open Government Licence v.3.0.


\bibitem{os_codepoint} \author{Ordnance Survey} Code-Point with Polygons. Available at: https://www.ordnancesurvey.co.uk/products/code-point-polygons (Accessed: 23 July 2024). © Crown copyright and database rights 2024 Ordnance Survey (AC0000851941).

\bibitem{office_for_national_statistics_2011} Office for National Statistics. 2011 rural/urban classification. Available at: \url{https://www.ons.gov.uk/methodology/geography/geographicalproducts/ruralurbanclassifications/2011ruralurbanclassification} (Accessed 25 April 2024).
Licensed under the Open Government Licence v.3.0.
\end{thebibliography}
\end{document}